\documentclass [10pt]{article}
\usepackage{amsmath,amssymb,cite}
\usepackage[english]{babel}
\usepackage[dvips]{graphicx}
\usepackage{graphicx}
\usepackage{indentfirst}
\numberwithin{equation}{section}
\begin{document}
\begin{center}
{\large{\bf Comments on  the Entanglement Entropy on Fuzzy Spaces}}\\
\bigskip
Djamel Dou

 {\it Dept of Physics and Astronomy, College of Science, King Saud
University, P.O. Box 2455 Riyadh 11451,
Saudi Arabia.}\\

\end{center}

\begin{abstract}
We locate the relevant degrees of freedom for the entanglement
entropy on some 2+1 fuzzy  models. It is found that the entropy is
stored in the near boundary degrees of freedom. We give a simple
analytical derivation for the area law using $1/N$ like expansion
when only the near boundary degrees of freedom are incorporated.
Numerical and qualitative evidences for the validity of near
boundary approximation are finally given .
\end{abstract}

\section{Introduction }

The entanglement entropy provides a natural and quantum
statistical interpretation for the area scaling law
\cite{s1,sd,w,su}. Although it is not necessary that the B.H
entropy is the entanglement of vacuum fluctuations of quantum
fields in nature, the latter must be present and  any consistent
quantum theory of spacetime must account for them. On the other
hand, it is well known that the entanglement entropy is divergent
in ordinary quantum field theory due to the absence of an UV
cutoff. The need to UV cutoff and the finiteness of black hole
entropy are widely viewed as a direct manifestation for a discrete
nature underlying spacetime at the Planck scale, and points out to
a necessary reduction of the number of degrees of freedom on the
horizon \cite{s2,suh}. Indeed, the combination of quantum
mechanics with gravity leads undoubtedly to a fuzzy picture of
spacetime. A possible realization of this picture is offered by
non-commutative and fuzzy geometry. Recently the entanglement on
some fuzzy space models was computed and shown to be finite and
proportional to the degrees of freedom on the boundary, once the
length scale parameters are restored the area law is recovered
\cite{db}.

The questions that we want to address in this paper concerns the
location of the degrees of freedom (DF)\footnote{Throughout the
paper DF will stand for "degrees of freedom".} which give the
dominant contribution to the entropy and the validity of the near
boundary approximation. A a similar question regarding the
location of the relevant DF was addressed in \cite{ind1,ind2} in
the context of lattice regularization. The main finding in
\cite{ind1,ind2} was that the entropy is essentially dominated by
the entanglement between DF very close to the horizon whereas DF
far from have very small contribution. However, lattice and
momentum cut-off regularization although been useful in showing
the area law scaling of the entanglement entropy they break the
underlying symmetry of the space, and generally the leading
divergent coefficient is non-universal.  The search for the DF
relevant for the entanglement entropy on fuzzy spaces is of
interest. On one hand, the nonlocal character of these spaces may
render the DF far from the Horizon (the separating boundary) much
more relevant than in the commutative case. On the other hand,
this will bring in a subtle question concerning the choice
disjoint regions to define the entanglement entropy. Such question
was tackled in \cite{db} using heuristic arguments and confirmed
by the results obtained. Moreover, fuzzy regularization being
symmetry preserving offers better control on the DF and allow for
many analytical considerations.

 In section 2 and 3 we will discuss the relevant DF for the entanglement entropy for different
fuzzy 2+1 models, the results in some cases are compared to the
lattice regularization. It is found that on fuzzy spaces too the
entanglement entropy is dominated  by the DF near to the
separating boundary, and as far as this point is concerned the
non-commutativity and non-locality of fuzzy spaces have not
altered the picture obtained using lattice regularization. In
section 3 we give an analytical derivation for the area law in the
cases where
 the DF incorporated  become infinitesimally close to the boundary
in the macroscopic limit. The results are derived using $1/N$ like
expansion.It is found that the area law is almost dictated by the
forms of the fuzzy potentials and the general properties of the
entanglement entropy. Finally we give qualitative arguments and
numerical evidences for the fact that the near horizon
approximation of the theory is enough to capture the entanglement
entropy in the macroscopic limit.

 \section{Fuzzy Sphere vs Continuum Sphere}
 We start by giving a brief account of the main formalism and results obtained in
 \cite{db}.

Consider a real scalar theory on $\bf{R}\times \bf{S}^2$, where
$\bf{S}^2$ is a sphere of radius $R$. The Hamiltonian  after
regularization in the cylindric coordinates is given by

\begin{equation}\label{11}
    H=\sum_{m=-\infty}^{+\infty}H_m
\end{equation}

\begin{equation}\label{10}
    H_m =\sum_{A,B=1}^{2N-1}\bigg[\delta_{A,B} \pi^A \pi^B+
    V^{(m)}_{AB}Q_m^A Q_m^B\bigg]
\end{equation}
 where
 \begin{eqnarray}
V_{AB}^{(m)}&=&\big(2-\frac{(N-A)^2}{N^2}-\frac{(N-A+1)^2}{N^2}+a^2{\mu}^2+\frac
{m^2}{N^2-(N-A)^2}\big){\delta}_{A,B}\nonumber\\
&-&(1-\frac{(N-A)^2}{N^2}){\delta}_{B,A-1}
-(1-\frac{(N-B)^{2}}{N^2}){\delta}_{A,B-1}
\end{eqnarray}

$ R=Na$ and $a$ is the lattice spacing (UV cutoff)\footnote{ In
this regularization the $z-$axis is replaced by a one-dimensional
lattice, i.e $z{\longrightarrow}z_n=n a$}.

If we now consider the DF residing in the upper hemisphere
unaccessible we construct the reduced density operator for the
ground state by integrating all the modes $ Q_m^{\alpha}$ for
$\alpha= N,...,2N-1$ corresponding to positive $z$ for all values
of $m$.

The resulting entanglement was computed numerically and found to
be

\begin{equation}\label{}
    S_N= 0.465 N=\frac{0.465}{2\pi a} A
\end{equation}
where $A =2\pi R$, which is the area  law in $2+1$ dimension.

Consider now instead  a free scalar field on ${\bf R}{\times}{\bf
S}^2_N$, where ${\bf S^2}_N$  is a fuzzy sphere of  matrix
dimension $N=2l+1$. The action reads

\begin{eqnarray}
S_N=\frac{1}{N}\int dt L~,~L=\frac{1}{2}Tr \bigg(\dot{\phi}^2
-\phi\big[{\cal L}_i^2+{\mu}^2\big]\phi \bigg).\label{2.1}
\end{eqnarray}
The scalar field $\phi$ is an $N{\times}N$ hermitian matrix with
mass parameter $\mu$. The Laplacian ${\cal L}_i^2$ is the $ SU(2)$
Casimir operator given by ${\cal L}_i^2={\cal L}_1^2+{\cal
L}_2^2+{\cal L}_3^2$ with action defined by ${\cal
L}_i({\phi})=[L_i,\phi]$ and ${\cal
L}_i^2({\phi})=[L_i,[L_i,\phi]]$. The $L_i$ satisfy
$[{L}_i,{L}_j]=i{\epsilon}_{ijk}{L}_k$ and they generate the
$SU(2)$ irreducible representation of spin $l=\frac{N-1}{2}$.

It turns out that the Hamiltonian of this action is better
expressed in terms of new variables $ Q^{(m)}_a$ related to matrix
elements of  $ \phi$ by introducing a convenient parameterization
as follow ,
$$
Q^{(m)}_a =\Phi_{a,a+m}  ~~~\text{for}~ m \geq 0~~,~~Q^{(-m)}_a
=\Phi_{a+m,a+m}   ~~~\text{for} ~m \leq 0~, ~~ \text{where}~~
\Phi= Re \phi +Im \phi
$$

 Using these new field variables the Hamiltonian  is
brought into the following form,
\begin{eqnarray}
H=\sum_{m=-(N-1)}^{N-1}
H_m=\sum_{m=-(N-1)}^{N-1}\sum_{a,b=1}^{N-|m|}\bigg[\frac{1}{2}(\pi^{(m)}_a)^2+\frac{1}{2}V_{ab}^{(m)}Q^{(m)}_a
Q^{(m)}_b\bigg].\label{2.2}
\end{eqnarray}
where
\begin{eqnarray}\label{Vfs}
V_{ab}^{(m)}=2\bigg[\big(c_2+\frac{{\mu}^2}{2}-A_aA_{a+|m|}\big){\delta}_{a,b}
-\frac{1}{2}B_{a-1}B_{a-1+|m|}{\delta}_{a-1,b}-\frac{1}{2}B_{a}B_{a+|m|}{\delta}
_{a+1,b}\bigg].\label{V}
\end{eqnarray}
where $\pi^{(m)}_a = \dot{Q}^{(m)}_a$ and $ B_a = \sqrt{a(N-a)}$
and $ A_a =-a+\frac{N+1}{2}$.

With this result one can see that the free theory splits into
$2(2l)+1$ independent sectors $ \{\mathcal{H}_m\} ,
m=-(N-1),\cdot\cdot\cdot\cdot, (N-1)$, each sector $\mathcal{H}_m$
has $N-|m|$ degrees of freedom ($N-|m|$ coupled harmonic
oscillator) and described by a Hamiltonian $H_m$.   Take now each
sector ${\cal H}_m $ and trace over half of the degrees of
freedom. For a fixed $N$ and $m$ the number of degrees of freedom
in the sector ${\cal H}_m $ is $N-|m|$, if $N-|m|$ is even we
trace out the following degrees of freedom
\begin{eqnarray}
Q^{(m)}_{1},Q^{(m)}_{2},\cdot\cdot\cdot,Q^{(m)}_{k},
k=\frac{N-|m|}{2}
\end{eqnarray}
if $N-|m|$ is odd we have two options, either we trace out
\begin{eqnarray}
Q^{(m)}_{1},Q^{(m)}_{2},\cdot\cdot\cdot,Q^{(m)}_{k}, k=
\frac{N-|m|-1}{2}
\end{eqnarray}
or we trace out
\begin{eqnarray}
Q^{(m)}_{1},Q^{(m)}_{2},\cdot\cdot\cdot, ,Q^{(m)}_{k},k=
\frac{N-|m|+1}{2}
\end{eqnarray}
However both options lead to the same entanglement entropy for
large $N$ and the degrees of freedom $
Q^{(m)}_{\frac{N-|m|+1}{2}}$ will be interpreted as boundary
degrees of freedom and there are $N$ of them.
 This corresponds in the original matrix notation
to dividing the matrix $\phi$ into two parts, left upper
triangular matrix $ \phi_{U}$ and  right lower triangular one
$\phi_{L}$ . For example, for $N=5$ the $\phi_{U}$ and $\phi_{L}$
will look as follows

\begin{eqnarray}
\phi_{U}=\left (\begin{array}{ccccc}
   {\phi}_{11} & {\phi}_{12} & {\phi}_{13} & {\phi}_{14} & 0 \\
   {\phi}_{21} & {\phi}_{22} & {\phi}_{23} & 0 & 0 \\
   {\phi}_{31} & {\phi}_{32} & 0 & 0 & 0 \\
   {\phi}_{41} & 0 & 0 & 0 & 0 \\
   0 & 0 & 0 & 0 & 0
 \end{array}\right)~,~
\phi_{L}=\left (\begin{array}{ccccc}
  0& 0 & 0 & 0 & \phi_{15}\\
    0 &0 & 0 & \phi_{24}  &  \phi_{25}\\
      0 & 0 & \phi_{33}  & \phi_{34} & \phi_{35}  \\
        0 &  \phi_{42} & \phi_{43} & \phi_{44}  & \phi_{45} \\
      \phi_{51}& \phi_{52} & \phi_{53} & \phi_{54} &  \phi_{55}
\end{array}\right)
\end{eqnarray}

The components $ \phi_{51}, \phi_{42}, \phi_{33},
\phi_{24},\phi_{15}$ are the boundary degrees of freedom.
$\phi_{U}$ and $\phi_{L}$ can be given the interpretation of
corresponding to functions with disjoint supports, one on the
lower half and the other on the upper half of the fuzzy sphere.

The  reduced density operator is given by
\begin{equation}\label{2.3}
{\rho}_{\rm red} = \bigotimes_{m =-(N-1)}^{N-1} {\rho}_{\rm
red}^{(m)}
\end{equation}
and the associated entropy  is
\begin{equation}\label{2.4}
    S_N= S_0 +2 \sum_{m=1}^{2l}S_m , ~~ N=2l+1.
\end{equation}

\begin{equation}\label{5}
    S_{ m}= \sum_i\bigg[
\log\big(\frac{1}{2}\sqrt{{\lambda^m}_i}\big)+
    \sqrt{1+{\lambda^m}_i}\log\bigg(\frac{1}{\sqrt{{\lambda^m}_i}}+
    \sqrt{1+\frac{1}{{\lambda^m}_i}}\bigg)\bigg]
\end{equation}
Where ${\lambda^m}_i$ are  the eigenvalues of the following matrix
\begin{equation}\label{6}
    \Lambda_{i,j}^{(m)}= -\sum_{\alpha}W_{i\alpha}^{(m)-1}W^{(m)}_{\alpha j}
\end{equation}

$W^{(m)}$ is the square root matrix of $V^{(m)}$  and $ W^{(m)-1}$
is the inverse of $W^{(m)}$. The indices $i,j$ run over the
available region and $\alpha$ over the region to be traced out.

The resulting entanglement entropy can be computed numerically and is found to be equal to ( for large $l$ or $N$),
  \begin{equation}\label{2.6}
    S_N= 0.39 \sqrt{l(l+1)}=  0.39 \frac{R}{\theta}= \frac{0.39 A}{2\pi\theta}
\end{equation}
 Equation (\ref{2.6}) can also be given another interesting interpretation, namely the
entropy is proportional to the number of boundary DF as
$$
S_N= 0.19(2l+1)
$$
We are now  ready to address the question regarding the DF which
give the dominant contribution and to what extent the DF far from
the boundary contribute to the entropy.

To that end we shall perform the following operation on the
matrices $V^{(m)}$.

\begin{figure}[ht]
\begin{center}
\includegraphics[width=9cm,angle=0]{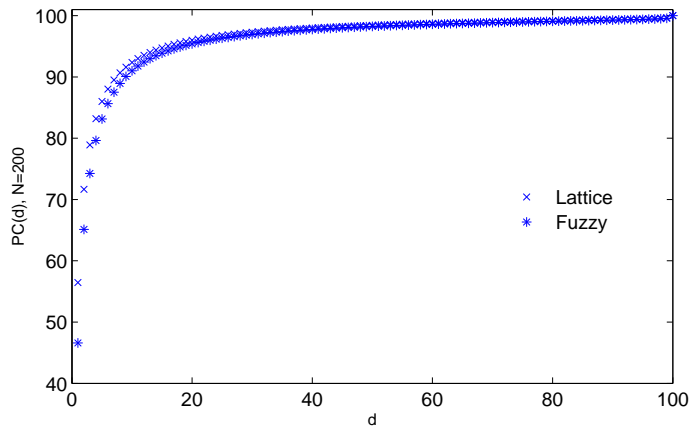}
\caption{{  The percentage contributions to the entire
 entanglement entropy as function of the number of the
incorporated DF for the lattice and  fuzzy regularization of the
sphere. }}
\end{center}
\end{figure}

The  operation in question  amounts to switching off the
entanglement between the DF which are at distance less than
$d$-lattice spacing \footnote{Note here that we loosely speaking
use the term lattice spacing for the fuzzy regularization too.}
from the separating boundary and the remaining (i.e we include
only $d$ DF from the outer and an equal number from the inner
region). This is achieved by setting by band to zero all the
off-diagonal terms except those corresponding to the $2d+1$ DF to
be incorporated .This operation is the same as one of the
operations considered in \cite{ind1,ind2}.

 We compute $S_N(d)$ and vary $d$, including successively more
distant DF until all DF are included. Each time we compute the
percentage contribution to the entropy as a function of $d$
\begin{equation}\label{}
  PC(d)=100\times S_N(d)/S_N
\end{equation}
The results for both models, the fuzzy and lattice regularization,
are depicted in Figure 1.

The first thing to note is that, except for the first few DF, the
relevance of DF far from the horizon for fuzzy sphere is exactly
the same as that in the lattice regularization, the
non-commutativity and non-locality have not brought anything new
to this point.
   When only the first few DF are included the lattice
and fuzzy regularization have slightly different responses.  For
the lattice regularization  $56\%$ and $71 \%$   of the entire
entropy is recovered at $d=1$ and $2$ respectively, whereas for
the fuzzy sphere we only reach $46\%$ and $65\%$. At $d=10$ we
recover $92\%$ in both the fuzzy and the lattice case. For $d >
10$ the two curves almost  fit together.
 This similarity between the lattice and the fuzzy regularization
regarding the DF dominating the entropy confirms and sharpens  the
heuristic argument given in \cite{db} to define the boundary in
fuzzy sphere case.

Now, the important question to ask here is whether in the
macroscopic limit the entropy will captured by the DF
infinitesimally close to the boundary. Indeed it is hard to settle
this question by numerical methods, one needs an analytical
estimation for the contributions of the DF residing at distances
which remain finite in the macroscopic limit. Nevertheless, our
numerical results suggest strongly that in the macroscopic limit
the entropy is given by the contributions of the DF which become
infinitesimally close to the boundary.  For example for $l=2000$
we find that essentially more than $99\%$ is stored at distance
less than $d=150$.

In the last section we will reconsider this point from analytical
point of view and give further numerical and qualitative evidences
for this.

Before moving to other models there is a technical point that
needs to be mentioned here. Although performing the previous
operation in the lattice regularization case is straight forward,
because the matrices $V^{(m)}$ in such case all have equal
dimensions (equal number of DF), for the fuzzy case different
sectors corresponding to different $|m|$ have different numbers of
DF. For a given $d$, sectors with $N-|m| < d$ already exhaust
their maximal contribution, however,  we shall show later that in
the macroscopic limit these sectors give a negligible contribution
to the entropy.

\section{ Fuzzy Disc and Moyal Plan}

Let us now consider the fuzzy disc model  and perform the same
operations.

The model in question is defined as follow.Consider  a scalar
theory on ${R}{\times}{R}_{\theta}^2$ where ${ R}_{\theta}^2$ is
now the Moyal plane. The action is given by :
\begin{equation}
 S=\frac{1}{2}\int dt Tr ( \dot{\phi}^{2}-\phi\big(
\nabla^{2}+{\mu}^2)\phi).\label{disc}
\end{equation}
The trace is infinite dimensional and the Laplacian is given in
terms of creation and annihilation operators $a$ and $a^{+}$ by
the expression
\begin{equation}
\nabla^{2}\phi:=
-\frac{4}{\theta^{2}}[a^{+},[a,\phi]]=-\frac{4}{\theta^{2}}[a,[a^{+},\phi]].
\end{equation}
Let us recall that $[a,a^{+}]={\theta}$, where $\theta$ is the
noncommutativity parameter. The fuzzy disc is obtained from the
plane following  \cite{liz} as follows. We consider finite
dimensional $N{\times}N$ matrices ${\phi}$, viz
\begin{equation}
\phi=\sum_{n,m=0}^{N-1}\phi_{mn}\mid
m><n\mid~,~{\phi}^{+}={\phi}~,~{\phi}_{nm}^{*}={\phi}_{mn}.
\end{equation}
Then it can be shown that the Laplacian $\nabla^{2}$ acts on a
 finite  dimensional space of dimension $(N+1)^2$, i.e $\nabla^{2}\phi$
is an $(N+1){\times}(N+1)$ matrix. The action on ${\bf
R}{\times}{\bf D}_N^2$ is thus given by (\ref{disc}) where the
trace $Tr$ is simply cut-off at $N$. We denote this trace by
$Tr_{N}$. The radius of the disc is given by
\begin{eqnarray}
R^2=N{\theta}.
\end{eqnarray}

By introducing new variables similar to the ones introduced in the
fuzzy sphere case one obtains the following Hamiltonian

\begin{eqnarray} H=\sum_{m=-(N-1)}^{N-1}
H_m=\sum_{m=-(N-1)}^{N-1}\sum_{a,b=1}^{N-|m|}\bigg[\frac{1}{2}(\pi^{(m)}_a)^2+\frac{1}{2}VQ^{(m)}_a
Q^{(m)}_b\bigg].\label{HD}
\end{eqnarray}

where $V_{ab}^{(m)}$ is now given by

\begin{eqnarray}
V_{ab}^{(m)}&=&2\bigg[\big(2a+|m|-1+\frac{{\mu}^2{\theta}}{4}\big){\delta}_{a,b}
-\sqrt{(a-1)(a-1+|m|)}{\delta}_{a-1,b}-\sqrt{a(a+|m|)}{\delta}_{a+1,b}\bigg].\nonumber\\
\end{eqnarray}

\begin{figure}[ht]
\begin{center}
\includegraphics[width=9cm,angle=0]{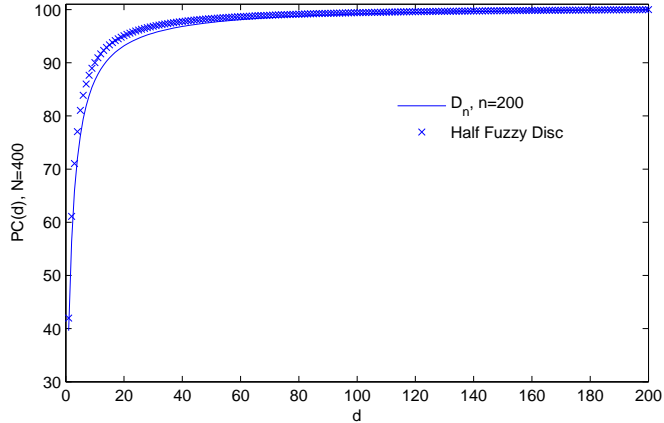}
\caption{{  The percentage contributions to  the entire
entanglement entropy as function of the number of the incorporated
DF for the two fuzzy disc cases.}}
\end{center}
\end{figure}

In the fuzzy disc model it turns out that there are two
interesting cases to consider for the entanglement entropy. The
first one results from tracing out the DF residing inside a
smaller subdisc $D_n \subset D_N$; the second one is to trace half
of the fuzzy disc. The tow cases have different behavior if we
consider the Moyal plan limit. Whereas in the first case the
smaller disc will remain intact and finite when $N$ is sent to
infinity, in the second case  the ignored region blows up in the
Moyal plan limit.

The entanglement entropy resulting from tracing out a fuzzy
subdisc $D_n$ was computed numerically in \cite{db} and shown to
be given by

\begin{equation}\label{efd}
  S_N(n)=S_0+2\sum_{m=1}^{n} S_m = 0.23(2n+1)
\end{equation}

For the second case where half of fuzzy disc is traced out the
entanglement entropy turned out to be


\begin{eqnarray}\label{ehfd}
S_N= S_{0}+2\sum_{m=1}^{N-1} S_m= 0.34N
\end{eqnarray}

In both cases the entanglement entropy is proportional to the
boundary degrees of freedom and can be as well interpreted  as
proportional to the area of the separating boundary ( being here
the circumference of a  circle in the first case and the diameter
of the disc in the second one.)

Let us now address the same question we addressed for the fuzzy
sphere case and locate the DF which contribute most to the
entanglement entropy.

In the first case, where we trace out a subdisc $D_n$, we compute
the entanglement entropy $S_N(n,d)$ obtained by incorporating only
DF from the subdisc $D_n$ with $d$ lattice spacing from the
separating boundary, that is  DF residing in $D_n-D_{n-d}$,
similarly from the outer region we include only DF with $d$
lattice spacing from the boundary, i.e DF in $D_{n+d}-D_n$. Then
$d$ is run from $1$ to $n $ \footnote{Here we to avoid unnecessary
computational complications we have taken $n=N/2$.}.

For the second case the same operation is applied, which is very
similar to the fuzzy sphere case. The results for both cases are
depicted in Figure 2. Again we see that by including the first few
degrees of freedom from the boundary $98\%$ from the total entropy
is captured. Nevertheless DF far from the boundary have small
contribute, this contribution become smaller and smaller as we
move further from the boundary, the DF at $d=n$ contribute by less
than $10^{-2}\%$.

\section{On the Area Law and Near Boundary Approximation}

In the previous sections we showed that the vacuum fluctuations in
the vicinity of the horizon ( the separating boundary) are
responsible for giving the dominant to the entanglement entropy in
different fuzzy space models. The questions that we want
 address in this section concern the area law itself and the validity of the near
  boundary approximation in the macroscopic limit.

  The considerations of the previews sections suggest strongly that the entanglement entropy should
  be given by the entanglement between
  the inner and the outer DF residing at distances which become infinitesimally close to
  the boundary in the macroscopic limit. Therefore we shall start by giving a simple analytical derivation of the area law
  using $1/N$ like expansion\footnote{ $N$ may
stand here for $l$ , $n$ or $N$ depending on the model we are
considering.} in the case where the DF incorporated are
infinitesimally close to the boundary in large $N$ limit.

Consider first the fuzzy sphere model in the limit of very large
$l$ or $N$ and $d/l \ll 1$, .

As already mentioned for a fixed $d$ some sectors already exhaust
their maximal contribution, hence we shall distinguish two classes
of sectors.

The first class corresponds to the sectors with  $  |m| \geq
2l-2d+rem(m,2)\equiv m_d ~~$\footnote{ $rem(m,2)$ stands for the
reminder of $|m|/2$.}. These are the ones which exhaust their
maximal contribution to the entropy, therefore their analysis is
valid even when the entire number of DF is incorporated.

The second class corresponds to the sectors  with $ |m| <
2l-2d+rem(m,2)$.

Thus the entropy   naturally splits into two contributions
\begin{equation}\label{ss}
S_N(d) = S_{\prec} +S_{\succ}\;\;,\;\; S_{\prec}=
S_0+2\sum_{m=1}^{m_d-1} S_m(d,l)\;\;,\;\;
S_{\succ}=\sum_{m=m_d}^{2l} S_m(d,l)
\end{equation}
 We start by evaluating $ S_{\succ}$ and showing that it vanishes in
the $l\rightarrow \infty$.

From equation (\ref{Vfs}) it is not difficult to see that  for
$d/l \ll 1 $ the diagonal elements of $V^{(m)}$ dominate over the
off-diagonal ones, this allows for a perturbative evaluation of
$S_m$  . It is found that

\begin{eqnarray}\label{psm}
S_m=\frac{\lambda_m}{4} (1-\log \lambda_m/4) +O(\lambda^2),~~~
  \lambda_m= \frac{1-(\frac{m}{2l})^2}{1+(\frac{m}{2l})^2}
  <\frac{d^2}{4l^2}.
\end{eqnarray}

 from which it follows that

\begin{equation}\label{s1}
S_{\succ} < \frac{d^3}{4l^2}[ 1-2\log(d/4l)] +O (
  d^5\log(d/l)/l^4)
\end{equation}

this shows that $S_{\succ}$ vanishes in the limit $l \rightarrow
\infty$ for all $d < l^{2/3-\epsilon},~~\epsilon >0$. It follows
that sectors with number of DF less than $ l^{2/3-\epsilon}$ are
weakly entangled and have vanishing contributions in the limit of
large $l$ and therefore irrelevant for the entanglement entropy.

We now turn our attention to $S_{\prec}$.

We start by noting that for $d \ll l $ the relevant submatrices
$V_d^{(m)}$ of the matrices $V^{(m)}$ are given by

\begin{equation}\label{vd}
{V_d^{(m)}}_{ab}=(l^2+m^2/4)\delta_{ab} -\frac{1}{2} (l^2-m^2/4)[
\delta_{a+1,b}+\delta_{a,b+1}] ~~,~~ a,b=1 \cdot \cdot \cdot,
2d+1-rem(m,2)
\end{equation}

 where we have neglected terms of the order of $ld$.

If we scale $V^{(m)}$ by $1/l^2$ and note that the eigenvalues of
the matrices $\Lambda^{(m)}$ are invariant under overall  scaling
of the matrices $V_d^{(m)}$, we conclude that $S_m$ will depend
only on the ratio $m^2/(4l^2)~~$. This means that in (\ref{vd}) we
have neglected effectively terms of the order of $d/l$. In effect,
as far as the entanglement  is concerned the terms which we have
neglected in (\ref{vd}) are much less relevant than their order of
magnitude may suggest. This will shortly be confirmed by numerical
results.

Now, in  contrast to the previous class of sectors, these sectors
are strongly entangled because the off-diagonal terms are
generally of the same order of magnitude as the diagonal ones.
Thus no perturbative evaluation of  $S_{\prec}$ is possible.
Nevertheless equations (\ref{vd}) and (\ref{psm} ) are enough to
establish the area law.

From equations (\ref{psm}) and (\ref{vd}) we see that the $S_m$'s
depend exclusively on the ratios $ \frac{m}{2l}$ and the entropy
is therefore given by

\begin{equation}\label{}
S_N(d)= S(0,d)+2\sum_{m=1}^{2l} S(m^2/4l^2, d)
\end{equation}

 In the large $l$ limit
$S_N(d)$ can be well approximated by integral

\begin{equation}\label{sd}
S_N(d) = 4l\int_0^1 S(x^2,d) dx
\end{equation}

 Equation (\ref{sd}) establishes the area law; and shows that for $ l \gg
d$ the percentage by which the near boundary DF contribute becomes
independent of the size of the boundary in the macroscopic limit.
Of course, equation (\ref{sd})  is understood to be exactly valid
only in the strict limit, otherwise  correction  terms
 would be present  and which vanish in the macroscopic limit. The above
analytical results can be confirmed by several numerical
evaluations of $S_N(d)$ for fixed $d$ and various values of $N$ or
$l$ . For example, if we define $c_N(d) = S_N(d)/l$ it is found
that $c_{1000}(d=20)-c_{1600}(d=20) \sim 10^{-6} $.

 Let us now consider the fuzzy disc model. In the case where we
trace out a subdisc $D_n$ and incorporate only  DF from the outer
and inner region with $d$ lattice spacing distance from the
boundary, we too find that we have
 to distinguish two classes of sectors.

The first class with $ |m| \ge n-d$. In this case we again observe
that for $ d \ll n$ the diagonal elements dominate over
off-diagonal ones and  $S_m $ can be evaluated perturbatively . We
find

\begin{equation}\label{ds}
S_m=\frac{\lambda_m}{4} (1-\log \lambda_m/4)
  +O(\lambda^2)~~ ,~~\lambda_m =\frac{1-|m|/n}{(2-|m|/n)^2}
\end{equation}

and it follows that

\begin{equation}\label{s2}
S_{\succ} < \frac{d^2}{8n}(1-\log(d/16n))+O(d^2\log(n/d)/n^2)
\end{equation}

Again we see that sectors with number of DF  negligible compared
to $ n$ give small and vanishing contributions   in the
macroscopic limit. However, the rate by which $S_{\succ}$
approaches zero is slower than  the fuzzy sphere case.

It should be noted  also that, unlike the fuzzy sphere case, not
all sectors with $ |m| \ge |m|-d$ exhaust  their maximal
contribution. This is due to the fact that for some sectors
certain   outer DF are still left out.

Consider now the second class of sectors, namely the ones with $
|m| < |m|-d+1$. For $d \ll n$ we find

\begin{equation}\label{v}
{V_d^{(m)}}_{ab}= (2n-|m|)\delta_{ab} -\sqrt {n(n-|m|)}[
\delta_{a+1,b}+\delta_{a,b+1}] +O(d)
\end{equation}

Again, we use the fact that $S_m$ is invariant under scaling of
$V^{(m)}$ and conclude that $S_m$  depends only the ratio $|m|/n$
and $d$. Therefore

\begin{equation}\label{sdfd}
S_N(d) = S_{\succ}(|m|/n,d)+S_{\prec}(|m|/n,d)
\end{equation}

in the large $n$ limit (\ref{sdfd}) can be approximated by and
integral, which becomes  exact in the limit $n \rightarrow
\infty$,

\begin{equation}\label{isd}
S_N(d) =2n\int_0^1 S(x,d) dx ~~
\end{equation}

This establishes the area law and shows that percentage by which
the DF contribute to the entire  entropy is independent of $n$ in
the macroscopic limit. This result can as well be confirmed by
several numerical calculations.

The case when we trace out half of the fuzzy disc is technically
similar to the fuzzy sphere, however the asymptotic behaviors are
similar to the subdisc case.

 Some comments about the above results are in order. We have established that  there are generally two
 classes
 of sectors. The first class is what we may call the irrelevant
 sectors or the weakly entangled ones, such sectors  have a small number of DF \emph{compared} to the number of
 the boundary DF and have vanishing contribution in the macroscopic limit.
 The second class is made of sectors which have number of DF  of the order of the number of
 boundary DF. These sectors are strongly entangled and  give the essential contribution to   the entanglement entropy.
 However, these results go somehow
against to what one may have naively guessed. The fact the entropy
is proportional to the number of sectors does not mean that all
sectors have relatively comparable contributions. Indeed, it is
the fact that the eigenvalues of the reduced density operator
depend only on the ratios $ m^2/4l^2$ or $|m|/n$ which leads
naturally to the area law, and this in turn will be related below
to the validity of the near boundary approximation.

It is interesting to mention here that in deriving equations
(\ref{sd}) and (\ref{isd}) we have made no real explicit use of
the eigenvalues or $S_m$, the area law is almost dictated by the
general properties of the entanglement entropy and the form of the
fuzzy potentials $V^{(m)}$'s .

Note that quite similar remarks apply to the lattice
regularization of the continuum sphere; sectors with $m$ much
larger than $N$ have small vanishing contributions \cite{db}.
However in the lattice regularization no analytical derivation of
the area seems easy to obtain in the limit $d \ll N$.

We end this paper by reconsidering the area law for the
entanglement entropy when we include all DF. We shall do this for
the fuzzy sphere. Similar arguments hold for the fuzzy disc. The
aim is to qualitatively argue that in the limit of large $N$ or
$l$ equation (\ref{sd}) remains true when we include all DF from
the inner and the outer region, that is $S_m$ will be functions of
$m^2/4l^2$ only .

We start by noting that the approximation given in equation
(\ref{vd}) remains valid for all DF at $d$-lattice distance from
the boundary as long as $d/l \rightarrow 0$ in the macroscopic
limit. Also, it is easy to see that the same approximation is
valid for
 the weakly entangled sectors.
Now,  for the remaining DF from the strongly entangled sectors one
has to distinguish tow classes of DF. The ones at $d$ such that $
(l-m/2 -d )/l \ll 1$, these DF are the ones that are far away from
the boundary, and the ones for which $d$ is of the order of $l$,
these are at an intermediate distances. For the first class it is
not difficult to show that their corresponding off-diagonal
elements die-off like $1/l$ and therefore they disentangle from
the other DF and become irrelevant in the macroscopic limit.
  For the second class none of the
above approximations is valid. However our numerical calculations
show that those DF are as well irrelevant. A quantitative argument
for this can be given as follows.

Let $\lambda$ be one of the eigenvalues of  a given
$\Lambda^{(m)}$, equation (\ref{6}). Any eigenvalue will be a
function of the off-diagonal elements of the matrix
  $V^{(m)} $ \footnote{After scaling the matrices by $1/l^2$ the diagonal elements become irrelevant for
  the argument and the off diagonal elements becomes less than $1/2$.}, $(v_{12}, v_{23},\cdot \cdot \cdot v_{k,k+1}, \cdot \cdot \cdot
  v_{n-1,n})$,  where $v_{k,k+1}$ is the boundary element which couples the
  inner DF to the outer ones.  First, it is by default that all the
  eigenvalues vanish identically if $v_{k,k+1}$ is set to zero.
  Also, the eigenvalues are all decreasing functions of the
  off-diagonal terms. On the other hand, setting one of the off-diagonal
 elements  at  position $d$ to zero would kill the contributions of all the
  successors with positions $p>d$. This leads to the conclusion that the contribution of DF at position $d$
    should be a decreasing function of the product $\prod_{i<d} v_{k+i,k+i+1}$, hence the  contribution of this DF
    will be suppressed in view
   of the fact that the elements $ v_{i,i+1}$ are all less than $1/2$.
Indeed if we assume the existence of a Taylor expansion  for the
  eigenvalues in terms of the off-diagonal elements, it is easy to see
  that every term in the expansion which contains some power of an
  off-diagonal element located at position $d$ must be accompanied
  by some power of all the precedent off-diagonal elements.

     The above  argument is of course only suggestive.
Indeed it would be interesting to obtain an asymptotic estimation
of the contribution of a DF at distance of the order of $l$.

Now, as an evidence for the above arguments let us push the near
boundary (horizon) approximation far beyond its region of validity
and assume it to be valid everywhere and compute the resulting
entanglement entropy including all DF. That is, we take our
starting potentials the ones given by (\ref{vd}) instead of the
ones given by (\ref{Vfs}),

\begin{equation}\label{vn}
{V^{(m)}}_{ab}=(l^2+m^2/4)\delta_{ab} -\frac{1}{2} (l^2-m^2/4)[
\delta_{a+1,b}+\delta_{a,b+1}] ~~,~~ a,b=1,   \cdot \cdot \cdot
2l+1-m,
\end{equation}

 compute the entanglement entropy incorporating all DF and
compare it to the one obtained using the exact form of $V^{(m)}$
of equation (\ref{Vfs}). Note first that the area law follows
automatically from (\ref{vn}), as all eigenvalues are functions
only of $m^2/4l^2$  (after scaling all matrices by $1/l^2$).

 Table 1 shows the values of the scaled entropy $S_N/l$ obtained
using the extended near boundary approximation of the potentials
against the values obtained by the original (exact) ones.
According to these numerical results the near boundary
approximation capture almost the exact value and become more
accurate as $l$ is pushed towards larger values, the deviation
from the exact true value is of the order of $1/l$. In the
macroscopic limit we would expect to obtain an exact agreement.

Indeed, it is not difficult to see why using the near boundary
approximation every where gives the same result in the large $l$
limit. First, the near boundary approximation capture the true
contribution of all near boundary DF and is valid for the weakly
entangled sectors.  For the DF far from  the boundary, the
potentials given by (\ref{vn}) provide essentially the same order
of suppression for their contributions  as the exact potentials
do.

\begin{table}
  \centering
\begin{tabular}{|r|r|r|r|r|r|} \hline
$l$$~~~~~~~~$ &200&300&500&600&900\\ \hline
 Exact form$~~~~~~~~~~$ &0.3960&0.3960&0.3960&0.3960&0.3960\\\hline
 Near boundary approximation &0.3929&0.3939&0.3948&0.3950&0.3953\\ \hline
\end{tabular}
 \label{table}
\end{table}

The validity of the near boundary approximation shows that for the
entanglement entropy all that matters is the near boundary (
horizon) geometry. This result goes in agreement with  the
standard results and paradigm about  black hole thermodynamics
 and field theories in the presence of black hole \cite{su,suh,stro}.
 Finally, it is interesting to note that despite
the non-locality and UV-IR mixing phenomena on fuzzy and
non-commutative spaces, entanglement entropy is still controlled
by the near boundary geometry.

 \paragraph{Acknowledgements}
The author  would like to thank E.I Lashin for useful discussions.
This work  is supported by King Saud University, College of
Science-Research Center Project No: (phys/2008/34).
\bibliography{mybibliography}

\end{document}